\documentclass[12pt,preprint]{aastex}

\shorttitle{The WISE IR Excess of GALEX1931} 
\shortauthors{Debes et al.}
\begin{document}
\title{The WIRED Survey I: A Bright IR Excess Due to Dust Around the Heavily Polluted White Dwarf GALEX J193156.8+011745\footnote{Based on data gathered with the 6.5 meter Magellan Telescopes located at Las Campanas Observatory, Chile}}
\author{John H. Debes\altaffilmark{1,2}, D. W. Hoard \altaffilmark{3}, Mukremin Kilic\altaffilmark{4,5}, Stefanie Wachter \altaffilmark{6}, David T. Leisawitz\altaffilmark{1}, Martin Cohen\altaffilmark{6}, J.~Davy Kirkpatrick\altaffilmark{6}, Roger L. Griffith\altaffilmark{6}}

\altaffiltext{1}{Goddard Space Flight Center, Greenbelt, MD 20771}
\altaffiltext{2}{NASA Postdoctoral Program Fellow}
\altaffiltext{3}{Spitzer Science Center, California Institute of Technology, Pasadena, CA 91125}
\altaffiltext{4}{Smithsonian Astrophysical Observatory, Cambridge MA 02138}
\altaffiltext{5}{Spitzer Fellow}
\altaffiltext{6}{IPAC, California Institute of Technology, Pasadena, CA} 
\altaffiltext{7}{Monterey Institute for Research in Astronomy, Marina, CA 93933}

\begin{abstract}
With the launch of the Wide-Field Infrared Survey Explorer (WISE), a new era of detecting planetary debris around white dwarfs has begun with the WISE InfraRed Excesses around Degenerates (WIRED) Survey.  The WIRED Survey will be sensitive to substellar objects and dusty debris around white dwarfs out to distances exceeding 100~pc, well beyond the completeness level of local WDs and covering a large fraction of known WDs detected with the SDSS DR4 white dwarf catalogue.  In this Letter we report an initial result of the WIRED survey, the detection of the heavily polluted hydrogen white dwarf (spectral type DAZ) GALEX J193156.8+011745 at 3.35 and 4.6 \micron.  We find that the excess is consistent with either a narrow dusty ring with an inner radius of 29~$R_{\rm WD}$, outer radius of 40~$R_{\rm WD}$, and a face-on inclination, or a disk with an inclination of 70$^\circ$, an inner radius of 23~$R_{\rm WD}$, and an outer radius of 80~$R_{\rm WD}$.  We also report initial optical spectroscopic monitoring of several metal lines present in the photosphere and find no variability in the line strengths or radial velocities of the lines.  We rule out all but planetary mass companions to GALEX1931 out to 0.5~AU.
\end{abstract}

\keywords{circumstellar matter--planetary systems--white dwarfs}

\section{Introduction}
The Wide-field Infrared
Survey Explorer (WISE) is a NASA medium class Explorer mission that
was launched on 14 Dec 2009 \citep{wright10}.  WISE mapped the entire
sky simultaneously in four infrared (IR) bands centered at 3.4, 4.6,
12, and 22~$\mu$m ({\it W1}, {\em W2}, {\em W3}, and {\em W4},
respectively) with $5\sigma$ point source sensitivities of
approximately 0.08, 0.1, 1, and 6~mJy, respectively.  WISE has several main goals, namely to take a census of cool stars and brown dwarfs close to the sun, probe the dustiest galaxies in the universe, and catalogue the Near Earth Object population \citep{wright10}.  The WISE mission will also provide crucial information about a diverse range of phenomena in the IR sky at a sensitivity 100 times better than IRAS.  

The WISE InfraRed Excesses around Degenerates (WIRED) survey is designed to detect infrared excesses around white dwarfs (WDs) using photometry from the WISE catalogue.  Dust, low mass companions, and cyclotron radiation from accreting magnetic WDs all emit at mid-IR wavelengths, providing a rich variety of sources to be discovered.  There are over 2131 WDs spectroscopically identified in the  McCook \& Sion catalogue \citep{mcook99,hoard07} and over 18,000 identified in the DR7 WD catalogue \citep{kleinman}--many of these objects will be detected with WISE.

Dusty WDs in particular provide information on the future fate of our own Solar System, as well as planetary systems around other stars.  Planetary systems can survive post main sequence evolution and mass loss as a central star becomes a WD \citep{duncan98}, though many objects in the inner system are expected to be destroyed through engulfment or evaporation \citep{villaver07,villaver09,nordhaus10}.  Rocky planetesimals can survive gas drag and sublimation during post-main sequence evolution \citep{jura08,dong10,bonsor10}, while simple models suggest icy planetesimals should be evaporated out to a few hundred AU for most stars \citep{stern90}.  The parent bodies of debris disks that have been observed with various {\em Spitzer Space Telescope} \citep{werner04} surveys evolve through post main sequence evolution and can become detectable during the planetary nebula phase \citep{chu07,chu09}, but rapidly become too cold to be observed at any wavelength \citep{bonsor10}.  

These planetesimals may once again become detectable later in a WD's evolution as it cools.  Roughly 25-30\% of cool isolated WDs show metal enrichment through optical spectroscopic detection of Ca or other lines \citep{zuckerman03,koester05,zuckerman10} or show emission due to heated gas \citep{gaensicke06,gaensicke07,gaensicke08}.  Metal polluted, hydrogen atmosphere WDs have short settling times for metals \citep{koester09} and are inferred to be actively accreting metal rich material.  Eighteen of the known metal enriched WDs show IR excesses due to optically thick dust located at a radius of $\sim$10~R$_{WD}$ \citep{jura03,reach05,kilic06,vonhippel07,farihi10}.

Given the short lifetimes of dust due to collisions or Poynting-Robertson drag at such distances, the optically thick disk of dust evolves on a viscous timescale, while the
presence of strong silicate emission features in most of these dusty
disks that have been observed with the Infrared Spectrograph on {\it Spitzer} points to an additional reservoir of optically thin material \citep{jura07b,reach05,reach09}.

The presence of dust within a region that should be devoid of any material due to post main sequence evolution of the WD progenitor is challenging to explain.  Planetesimals must be perturbed (presumably by a planet) on timescales that range from a few Myr to a few Gyr and tidally disrupted by the central WD \citep{jura03,jura08}.  The perturbation of planetesimals by the post main sequence
destabilization of giant planetary systems has been proposed, but to date no quantitative predictions for the lifetime or efficiency of this mechanism have been made \citep{debes02}.  

The circumstantial evidence, however, is compelling.  Order of magnitude estimates for the expected number of metal polluted WDs from \citet{debes02} are consistent with what is observed and is similar both to the frequencies of giant planets in orbit around early type stars \citep{johnson} and the estimated frequencies of close-in Earth mass planets \citep{howard10}.  Post main sequence planetary systems provide an important complementary sample to main sequence planetary systems and will provide crucial compositional information on extrasolar planetesimals impossible with other observational techniques.

Yet several questions about dusty and metal
enriched WDs remain.  The lifetime of dust, the exact structure of the dusty disks, and their evolution are all highly uncertain.  The answers to such questions may come via large number statistics from which trends and correlations can be identified.

In this respect, the WIRED survey is uniquely positioned to provide a large number of dusty WDs to help answer these and other questions.  WISE's 5$\sigma$ sensitivity limits of 0.08, 0.1, 0.8, and 6 mJy in the {\it W1}, {\it W2}, {\it W3}, and W4  bands are low enough to detect dusty disks in {\it W1} and {\it W2}, such as that detected around G29-38, out to $\sim$140~pc, and its bright silicate feature out to $\sim$55~pc.  A WD with T$_{eff}\sim$10,000~K, the approximate mean $T_{eff}$ of the SDSS DR4 survey \citep{eisenstein06}, has a $g$ magnitude of $\sim$18 at 140~pc \citep{bergeron95}.  There are 1800 WDs with g$<$18 in the DR4 catalogue and $\sim$1\% of WDs with T$_{eff}\sim$10,000~K (or a cooling age of 600~Myr for a 0.6 M$_\odot$ WD) harbor dusty disks \citep{farihi10}, implying as many as 18 could be detected with the WIRED survey.  The DR7 WD catalogue will contain twice as many
objects as the DR4 catalogue; in concert with the McCook \& Sion
catalogue \citep{mcook99,kleinman}, this could mean as many as 60--70 new dusty
WDs will be discovered.

In this Letter we present an initial result from the WIRED survey: the
detection of a significant IR excess around GALEX J193156.8+011745 (hereafter GALEX1931).  This DAZ was recently discovered to have a large amount of metal rich material in its photosphere, and a possible near-IR excess \citep{vennes10}.  The nature of the excess was unclear, however, because H and K photometry were consistent with either a brown dwarf or a dusty disk \citep{vennes10}.  In \S \ref{sec:valid} we demonstrate that WISE photometry of known dusty WD disks is consistent with previous {\em Spitzer} observations.  In \S \ref{sec:model} we report the {\it W1} and {\it W2} flux densities of GALEX1931, rule out any source contamination, and construct plausible models for the excess we observe around GALEX1931, finding a good fit to the photometry if we assume an optically thick dust disk.  Finally, we present spectroscopic follow-up of GALEX1931 to verify whether any variability in accretion has occurred on timescales that are long compared to the implied settling time for metals and to place limits on possible companions.  We present our conclusions in \S \ref{sec:conc}.  In addition to this work, GALEX1931 has been detected in the mid-IR from the ground (Mellis et al. 2011, submitted).

\section{Confirmation of WISE {\it W1} and {\it W2} First-Pass photometric measurements}
\label{sec:valid}
The measurements and images we used from the WISE catalogue are co-adds of all available data for GALEX1931 using ``First-Pass'' processing that used on-orbit performance of WISE as part of the pipeline.  To help test this pipeline, we have considered the {\it Spitzer} Infrared Array Camera (IRAC) photometry of previously reported dusty WDs that are also detected in WISE.  Since most of these systems are consistent with $\sim$1000~K blackbody emission from dust, the peak of their emission occurs at $\sim$3-5\micron\ and represents a departure in this wavelength range from the spectral slope of Vega that is used to calculate the zeropoints of the WISE photometric system \citep{wright10}.  Even so, blackbody emission with a temperature of $\sim$1000~K requires corrections at only the $\sim$2\% level as given in Table 1 of \citet{wright10}.  

Several dusty WDs have been observed with the IRAC camera aboard {\em Spitzer}.  The IRAC 1 and 2 bands have central wavelengths and passbands ($\lambda_c$=3.6 and 4.5\micron, respectively) that are quite close to the WISE 1 and 2 bands of 3.35 and 4.6 \micron.  Published photometry of WD disks in the IRAC bands is listed in Table \ref{tab:phot}, compared to the equivalent photometry in the WISE bands.  This list is a selection from the full set of known dusty disks.  The average difference in photometric measures in {\it W1} and {\it W2} are -12$\pm$7\% and -4$\pm$9\%.  Based on this result, we see no significant offsets between the two bands within the uncertainties and expect that our photometry is accurate to the level reported for absolute WISE photometry. This, however, is a small sample and a full comparison of all known disks and WD photospheres along with comparisons in the W3 band will be included in the full WIRED survey results, to be published at a later date.

\section{A Dusty Debris Disk around GALEX1931}
\label{sec:model}
\subsection{Ruling out Significant Source Contamination}
A source centered on GALEX1931's position (as given in \citet{vennes10} is clearly detected with WISE at a S/N$\approx$30 in the 2 bands.  The WISE catalogue lists detections at {\it W3} (S/N$\sim$5) and W4 (S/N$\sim$3), but inspection of the WISE images at these bands shows either source confusion or noise, and we take these detections as upper limits to the true flux of GALEX1931.  Figure \ref{fig:f1} shows the WISE images in {\it W1} and {\it W2}. The WISE photometry of GALEX1931 is listed in Table \ref{tab:phot2}.  A bright source is located $\sim$16\arcsec\ to the southeast, but contamination from this source is negligible.  At this separation the wings of the W1 and W2 PSF from the source are $\sim$6\% of the peak flux in GALEX1931 and a small fraction of the peak of the PSF (See Figure 11 of \citet{wright10}).  Additionally, the WISE first pass data provide profile fit photometry which can deblend and remove any resolved contamination.
  
GALEX1931 resides within 10$^{\circ}$ of the galactic plane, where the density of sources is relatively high.  This raises the possibility that a contaminating source with a red spectrum could be present within the WISE PSF at the coordinates of GALEX1931. For the measured W1 and W2 magnitudes of GALEX1931, the predicted combined stellar and extragalactic source counts based on models of the infrared sky for {\it W1} and {\it W2} are 4000 and 3600 sources per square degree per magnitude, respectively \citep{wainscoat92, cohen93}.  Integrated over the WISE PSF, with FWHM$\approx6''$, this corresponds to a probability of $\approx0.03$ that a source as bright as GALEX1931 will be found in a given (randomly chosen) PSF in this region of the sky.  Thus, for a randomly chosen location, we expect essentially zero sources as bright as GALEX1931 to lie within a WISE PSF.  Conversely, if we pick a specific location because we do expect to find a source there (i.e., GALEX1931), then there is a high probability that the detected source is the one we are looking for.

This alone cannot rule out the presence of an additional (fainter) contaminating source within the WISE PSF. However, if GALEX1931 is contaminated by an object redder than itself, then, unless the contaminating object was exactly coincident with the position of GALEX1931, the centroid of the detected source would shift at longer wavelengths as the contaminating source dominated.  This is not observed, but to assess the possibility that there may be fainter sources that contaminate the photometry of GALEX1931, we obtained archival V-band images from the EFOSC camera that were used to estimate the V magnitude of GALEX1931 \citep{vennes10} and new images in Ks with the Palomar 200$^\prime\prime$ Wide Field Infared Camera (WIRC).  The WIRC image was constructed from 7 dithered images taken on 29 August 2010 with 10s integrations.  The images were sky subtracted and combined to create the final image shown in Figure \ref{fig:f1}.  

The EFOSC image shows one source at a separation of $\sim$2\farcs2 that is within the {\it W1} and {\it W2} PSF, denoted as source A.  Seeing conditions were poor for our WIRC image, but we were able to resolve A sufficiently from GALEX1931.  We then performed relative photometry to determine its V-Ks colors.  

In the V band image source A is well separated from the GALEX1931 PSF and does not contaminate GALEX1931's V measurement.  We therefore estimated the V magnitude of source A by measuring the counts in the EFOSC image relative to GALEX1931, which has V=14 \citep{vennes10}.  When this is performed we obtain V=18.5 for source A.  We estimate that the uncertainty in this value is no more than 10\%.  

In the Ks band image, source A is heavily contaminated by the seeing disk of GALEX1931, and photometry of GALEX1931 is also contaminated by source A.  This required the construction of an empirical PSF from bright sources within the 8\arcmin$\times$8\arcmin WIRC field of view in order to isolate each object's photometry.  We chose 8 bright sources and median combined them to subtract off the GALEX1931 PSF to perform aperture photometry on source A, and vice versa. From aperture photometry of Source A using different scalings to the empirical PSF, we have determined that 5\% errors in scaling (the level at which the subtraction of GALEX1931 is unnoticeable from the background) corresponds to 10\% errors in source A's photometry.  This dominates other uncertainties so we adopt this value as the uncertainty in source A's Ks magnitude.  

We then performed relative photometry of both source A and GALEX1931, using other sources in the WIRC image that are present in 2MASS \citep{skrutskie}.  Of the 8 bright sources we chose, six had reliable Ks photometry in 2MASS (2MASS J19315790+0117363, J19315456+0117174, J19315382+0117593, J19315292+0118238, J19315111+0116247, and J19315752+0116132).  We took the standard deviation of our measured magnitudes for GALEX1931 to estimate the uncertainty in its photometry.  Based on our aperture photometry relative to these sources in our WIRC image, we derive an uncontaminated magnitude for GALEX1931 of 14.68$\pm$0.05 and 16.1$\pm$0.1 for source A.  GALEX1931's revised magnitude is 0.23 mag dimmer than reported in the 2MASS source catalogue.

Source A is probably stellar.  Its V-K color is 2.4$\pm$0.1, consistent with a mid-K spectral type object.  Given its relative faintness compared to GALEX1931 at Ks, it is unlikely to be a significant source of contamination at the {\it W1} and {\it W2} bands.  At these longer wavelengths a K star is well into the Rayleigh-Jeans tail and its flux density is declining roughly as  $\lambda^2$, while GALEX1931's flux density is increasing from Ks to {\it W1} in Table \ref{tab:phot2}.  

Based on the low probability of a coincident source, good agreement between the WISE {\it W1} and {\it W2} positions of GALEX1931 relative to the 2MASS and GALEX positions of GALEX1931, plus source A's V-K color, there is little chance that the {\it W1} and {\it W2} detections are due to a source other than GALEX1931. 

\subsection{The Nature of GALEX1931's Excess}
We next compare the observed photometry of GALEX1931 with an expected WD photosphere.  We assume $T_{eff}$=20890~K, $\log$~g=7.9, and a distance of 55~pc as inferred by \citet{vennes10} through spectral fitting.  Using synthetic photometry of DA models we can compare the expected photometry of GALEX1931 at V, 2MASS J, H, WIRC Ks, {\it W1}, and {\it W2} \citep{bergeron95,holberg06}.  In order to predict the flux densities of the WD photosphere in the WISE bands, we extrapolate blackbody emission for  $T_{eff}$=20890~K to the WISE central wavelengths.  Experience with {\em Spitzer} photometry of WDs at high S/N shows that small deviations from blackbodies can occur, but are generally not large in magnitude for hotter WDs \citep{debes06,tremblay07}.

When we do this we find a small correction of 1\% must be made between the WD model paramters reported by\citet{vennes10} and the Bergeron models, equivalent to either a slightly larger radius for the WD or a smaller distance.  Regardless, we get a good fit to both the V and J band photometry of GALEX1931 and use that to determine the wavelength at which the excess starts.

Within the uncertainties of the 2MASS and WISE photometry, the excess starts beyond the Ks band with an 18$\sigma$ excess at {\it W1} and continues to the {\it W2} band with an excess that is 22$\sigma$\ above the photosphere.  Our new WIRC photometry shows that the excess at H and K reported by \citet{vennes10} is probably spurious.  Inspection of the 2MASS Ks image indeed shows a slight extension of the PSF at the position of source A and our Ks magnitude is 0.23 mag fainter, consistent with Source A contaminating GALEX1931's 2MASS photometry.

Several simple models can be fit to the data and since there are 2 significant measures of the excess photometry, we must restrict ourselves to models that require as few free parameters as possible.  None of
these models are necessarily correct, but they provide useful
examples of what could, and could not, be producing the IR
excess.

The simplest model is that of a blackbody emitting at a particular temperature and with a given surface area assuming it is at the same distance as GALEX1931.  A good fit is given by a 900~K blackbody with a surface area 800 times larger than the WD's, implying a radius for either a disk or a companion 30 times larger than GALEX1931.  This should conclusively rule out the possibility of a companion at the distance of GALEX1931 as the source of the excess, since this would imply a radius for the companion of $\sim$0.3~$R_\odot$, too large for a brown dwarf.  Conversely, it could mean that a 900~K object with R$\sim0.1R_\odot$ is located 3 times closer than GALEX1931, implying a T~dwarf at 18~pc from the Sun.  This is, again, an unlikely scenario, as already
discussed with regard to the source counts in the vicinity of GALEX1931, not to mention the additional requirement that the T-dwarf would have to be precisely aligned with the position of GALEX1931.  In any case, such a scenario is easily tested by additional observations:  a T~dwarf that close would most likely have high proper motion that could easily be detected with a second epoch of NIR images in the K band.

If we restrict ourselves to a face-on ($i$=0) optically thick disk following the procedure of  \citet{jura03}, the best fit to the excess at shorter wavelengths is given by a narrow ring with an inner radius of 29~$R_{\rm WD}$ and an outer radius of 40~$R_{\rm WD}$ (See Figure {\ref{fig:f2}).  In this case, the maximum temperature of the dust is 1100~K.  

In reality, the disk likely has non-zero inclination relative to our line of sight.  When we produce model disks with varying angles of inclination, the
data allow a maximum inclination of 70$^\circ$ (which is constrained by the
upper limit flux density in the {\it W3} band; see Figure \ref{fig:f2}).  The emission comes from an optically thick disk that extends from
23~R$_{\rm WD}$ out to 80~R$_{\rm WD}$, and the maximum temperature of the dust at the inner edge is 1350 K.  If the disk has a true inclination smaller than 70$^\circ$, then the disk's outer radius must be smaller.  

The radii can be compared within the context of physical processes, such as the sublimation radius of the dust and the tidal disruption radius of minor bodies \citep{jura03,jura08}.  These radii typically bound the extent of dusty WD disks and are pointed to as primary evidence that the origin of the dust originates from tidally disrupted material that is then accreting onto the host WD after dust sublimation.  By definition, we take the inner edges of the disk to correspond to the sublimation radius, since sublimation must compete with other grain removal mechanisms which are poorly known.  There exist some dusty WDs that show interior radii much larger than expected for pure dust sublimation, these objects may represent the late stages of dusty disk evolution \citep[e.g. G 166-58 and PG 1225-079][]{farihi07,farihi10}.  The tidal disruption radius for a WD is given by $R_{tidal}\sim(C_{\rm tide} \rho_{\rm WD}/\rho_{\rm ast})^{1/3} R_{\rm WD}\sim44 C_{\rm tide}$ R$_{\rm WD}$, assuming a compact asteroid with a bulk density of 3 g~cm$^{-3}$, but could extend to $\sim$63~C$_{\rm tide}$R$_{\rm WD}$ for a porous asteroid with a bulk density of 1~g~cm$^{-3}$ \citep{davidsson99,jura03}.  C$_{\rm tide}$ is a parameter of order unity that reflects the details of a particular disrupting body's rotation and composition.  The possibility exists that the disk we observe extends beyond the typical tidal disruption radius assumed for rocky asteroids.

\section{Spectroscopic monitoring of GALEX1931's Photospheric Metal Lines}

High resolution optical spectroscopy of GALEX1931 shows strong absorption features due to Mg, Ca, Fe, O, and Si, implying high rates of accretion for all of these atomic species \citep{vennes10}.  \citet{vennes10} also noted a super-solar abundance of O and a sub-solar abundance of Ca, using this as an argument for the presence of a substellar companion polluting GALEX1931 with a wind \citep[e.g.][]{debes06}.  

If the spectral lines are due to accretion from either a wind or dust, they may vary on short timescales, especially since GALEX1931's settling time should be on the order of a few days  \citep{koester09} and such variability has been claimed in the past \citep{vonhippel07,debes07}.  To investigate this possiblility and to look for new absorption features in this heavily polluted photosphere, we observed GALEX1931 with the MIKE spectrograph on the Clay Telescope on UT 16-17 June 2010, 8 July 2010, and 2-3 August 2010, for a total of 11800~s of integration time on the source.  We used a 0\farcs7$\times$5\arcsec slit which corresponded to a resolution of $\sim$40,000 at the Ca K line.  Th-Ar comparison lamp spectra were taken near in time to each spectrum of GALEX1931.  Bad weather degraded the first night of spectra, but most strong lines were still detected.

Our data were extracted and flatfielded using the MIKE reduction pipeline written by D. Kelson, with methodology
described in \citet{kelson00} and \citet{kelson03}.  
In order to check for variability among the 5 nights of
observation, we calculated the equivalent width (EW) of every spectral
feature in each spectrum that was detected with an EW $>$50 m\AA. To calculate the equivalent width, we chose a window around the line equivalent to $\pm$3 times the FWHM of the line as determined by a simple Gaussian fit.  We chose several different polynomial fits to the continuum and added any systematic uncertainty in the continuum fit to our uncertainties in the equivalent width calculation.  Table \ref{tab:eqw} shows the individual line equivalent width measurements for each of the nights, as well as an average of the barycentric corrected velocities of the lines (with the exception of the Mg II doublet at 4481\AA) based on gaussian fits to each detected line.

In order to determine whether the line strengths vary, we assume that the equivalent width of each line is equal to the mean of the observations.  We then calculate the $\chi_\nu$ value of each line for its departure from a constant value.  Given the small number of samples for each line, we require $\chi_\nu$=40, equaling a probability of $<10^{-3}$ that the line deviates from a constant value even if our uncertainties were underestimated by 50\%.  From the lines we consider in Table \ref{tab:eqw}, we find no significant variability over our observations.  Despite a very short expected settling time of $\sim$6 days for elements such as Ca \citep{koester09}, no change in the accretion rate occurred to the level of 5-20\%.

From these five epochs, we can also place upper limits on the mass of any companions that may be present in short orbits.  Given that our observations are separated by at most three months and with minimum separations of a day, we would be sensitive to radial velocity variations with orbital periods that range from a few hours to $\sim$6 months, or for GALEX1931 with M=0.57$M_\odot$, orbits with semi-major axes that range from 0.004~AU to 0.5~AU.  The standard deviation of the velocities reported in Table \ref{tab:eqw} is 230~m/s, consistent with the level of epoch to epoch precision of MIKE without an iodine cell \citep{guillem10}.  Using this as an upper limit to the velocity semi-amplitude ($K$) for any putative companions, one can compute the upper limit to companion masses between our limiting orbital semi-major axes, given by $M_{\rm pl}\sin{i}=M_\star^{1/2} K (r_{\rm orb} G)^{-1/2}$, where $r_{\rm orb}$ is the orbital semi-major axis assuming a circular orbit.  For all separations we rule out any companion more massive than $\sim$5 M$_{\rm Jup}/\sin{i}$.
  It is possible still to have relatively massive planetary or substellar companions present around GALEX1931, but they would have to be in almost face-on orbits to be undetectable.  Furthermore, the lack of any excess at J, H, and K rules out all but substellar mass objects at any separation.  
  
\section{Conclusions}
\label{sec:conc}
We have detected an infrared excess at 3.35 and 4.6\micron\ around the WD GALEX1931 and determined it to be either a very narrow ring of dust or a wider dust ring within the tidal disruption radius of the WD.  Spectroscopic follow-up of GALEX1931 shows no evidence for any close companions down to planetary mass, which lends support to a disk interpretation for the excess.  Additionally, we have shown that the rate of accretion from the disk onto the WD does not significantly change at the observed epochs, equivalent to $\sim$8 metal settling times for GALEX1931's photosphere.  GALEX1931 is most likely accreting in the steady state regime, and the accretion itself does not vary at the level of $\sim$10-20\% from month to month.

We detected GALEX1931 with WISE in the {\it W1} and {\it W2} bands at high signal-to-noise, implying that we will be sensitive to similar analogues out to $\sim$180~pc, assuming the WISE sensitivity limits quoted above.  GALEX1931 represents a bright example of a larger sample of dusty WDs that will be detected with the WIRED Survey.

\acknowledgements
This research was supported by an appointment to the NASA Postdoctoral Program at the Goddard Space Flight Center, administered by Oak Ridge Associated Universities through a contract with NASA.    This work is based on data obtained from: (a) the Wide-Field Infrared
Survey Explorer, which is a joint project of the University of
California, Los Angeles, and the Jet Propulsion Laboratory (JPL),
California Institute of Technology (Caltech), funded by the National
Aeronautics and Space Administration (NASA); (b) the Two Micron All
Sky Survey (2MASS), a joint project of the University of Massachusetts
and the Infrared Processing and Analysis Center (IPAC)/Caltech, funded
by NASA and the National Science Foundation; (d) the Hale Telescope,
Palomar Observatory, as a part a continuing collaboration between
Caltech, NASA/JPL, and Cornell University; (e) the 6.5 meter Magellan
Telescopes located at Las Campanas Observatory, Chile; (f) the ESO
Telescopes at the La Silla or Paranal Observatories; (g) the SIMBAD
database, operated at CDS, Strasbourg, France; and (h) the NASA/IPAC
Infrared Science Archive, which is operated by JPL, Caltech, under a
contract with NASA.  We wish to thank D. Steeghs for obtaining spectra of GALEX1931 on 7-8 July and N. Morrell for obtaining spectra of GALEX1931 on 2-3 August.  M.C. thanks NASA for supporting his participation
in this work through UCLA Sub-Award 1000-S-MA756 with a UCLA FAU 26311 to 
MIRA.

\bibliographystyle{apj}

\begin{deluxetable}{cccccc}
\tablecolumns{6}
\tablewidth{0pt}
\tablecaption{\label{tab:phot} WISE and IRAC photometry of Dusty WDs}
\tablehead{
\colhead{Name} & \colhead{{\it W1} ($\mu$Jy)} & \colhead{IRAC1 ($\mu$Jy)} & \colhead{{\it W2} ($\mu$Jy)} & \colhead{IRAC2 ($\mu$Jy)} & \colhead{Ref.} \\
}
\startdata
GD 40  &  245$\pm$10  &  230$\pm$12  & 213$\pm$18 & 199$\pm$10  & 1 \\
GD 16   & 390 $\pm$13 & 486$\pm$24 &   486$\pm$23  & 508$\pm$25 & 2 \\
GD 362  & 314$\pm$7   & 380$\pm$20 &   413$\pm$12   & 395$\pm$20 & 3 \\ 
HE 2221-1630  &  186$\pm$9  &  204$\pm$10 &  140$\pm$15  & 172$\pm$9  & 4 \\
\enddata
\tablerefs{1--\citet{jura07}, 2--\citet{farihi09}, 3--\citet{jura07b}, 4--\citet{farihi10}}
\end{deluxetable}

\begin{deluxetable}{cccccc}
\tablecolumns{6}
\tablewidth{0pt}
\tablecaption{\label{tab:phot2}K$_s$ and WISE photometry of GALEX1931}
\tablehead{
\colhead{Name} & \colhead{\it K$_s$ ($\mu$Jy)} & {\it W1} ($\mu$Jy) & {\it W2}  ($\mu$Jy) &  {\it W3} ($\mu$Jy) & W4 ($\mu$Jy) \\
}
\startdata
GALEX1931 & 895$\pm$45 & 1024$\pm$36 & 925$\pm$33 & 640\tablenotemark{a} & 2097\tablenotemark{a} \\
\enddata
\tablenotetext{a}{These flux densities are upper limits}
\end{deluxetable}

\begin{deluxetable}{ccccccc}
\tablecolumns{7}
\tablewidth{0pt}
\tablecaption{\label{tab:eqw} Equivalent widths of Selected Atomic Elements}
\tablehead{
\colhead{Element} & \colhead{Epoch 1} & \colhead{Epoch 2} & \colhead{Epoch 3} & \colhead{Epoch 4} & \colhead{Epoch 5} & \colhead{$\chi_{\nu,\rm const}$} \\  
 & \colhead{16 Jun} & \colhead{17 Jun} & \colhead{8 Jul} & \colhead{2 Aug} & \colhead{3 Aug} & 
}
\startdata
\multicolumn{7}{c}{Equivalent Widths (m\AA)} \\
Si (3856\AA) & 84$\pm$20  & 37$\pm$8  & 52$\pm$4  & 43$\pm$3  &45$\pm$6  & 15.2 \\
Ca (3933\AA) & 90$\pm$20  & 39$\pm$13  & 48$\pm$4 & 55$\pm$3 & 64$\pm$6   & 16.0\\
Si (4128\AA) & 80$\pm$30 & 54$\pm$11 & 57$\pm$6 & 58$\pm$4 & 63$\pm$8 & 2.5\\
Mg (4481\AA) & 410$\pm$40 & 390$\pm$20  & 380$\pm$10 & 370$\pm$30 & 410$\pm$20 & 3.1 \\
Si (5957\AA) & 30$\pm$19 & 59$\pm$13 & 47$\pm$12 & 56$\pm$7 & 50$\pm$11 & 2.8 \\
Si (5979\AA) & 65$\pm$25 & 31$\pm$6 & 46$\pm$6 & 58$\pm$4 & 48$\pm$10 & 13.8 \\
Si (6347\AA) & 132$\pm$24 & 130$\pm$10 & 135$\pm$7 & 128$\pm$8 & 136$\pm$12 & 1.1 \\
Si (6371\AA) & 128$\pm$16 & 110$\pm$10 & 101$\pm$7 & 100$\pm$6 & 77$\pm$10 & 10 \\
\hline
\hline
\multicolumn{7}{c}{V$_{\rm helio}$ (km s$^{-1}$)} \\
& 36.80 & 37.16 & 36.75 & 37.15 & 37.27 \\
\enddata
\end{deluxetable}

\begin{figure}
\epsscale{0.75}
\plotone{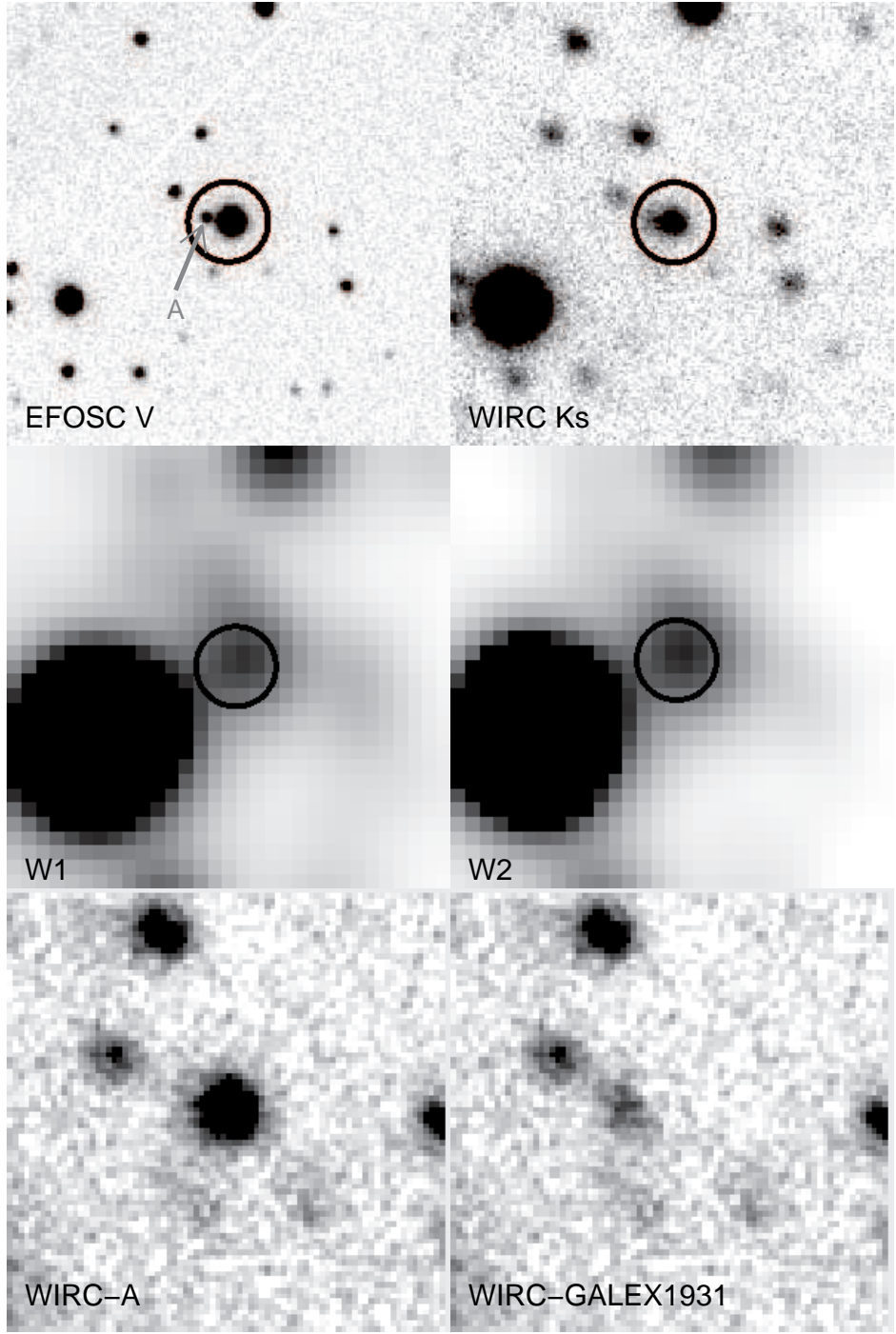}
\caption{\label{fig:f1} Images of GALEX1931 (top left) at V with the EFOSC camera on the NTT \citep{vennes10}, (top right) at Ks with the WIRC camera on the Palomar 200'', (middle left) {\it W1}, and (middle right) {\it W2}.  GALEX1931's position is centered in the circle in each panel and the potential contaminating source A is marked in the V band image.  The bottom two panels show the empirical PSF subtraction in Ks of (bottom left) source A and (bottom right) GALEX1931.}
\end{figure}

\begin{figure}
\plotone{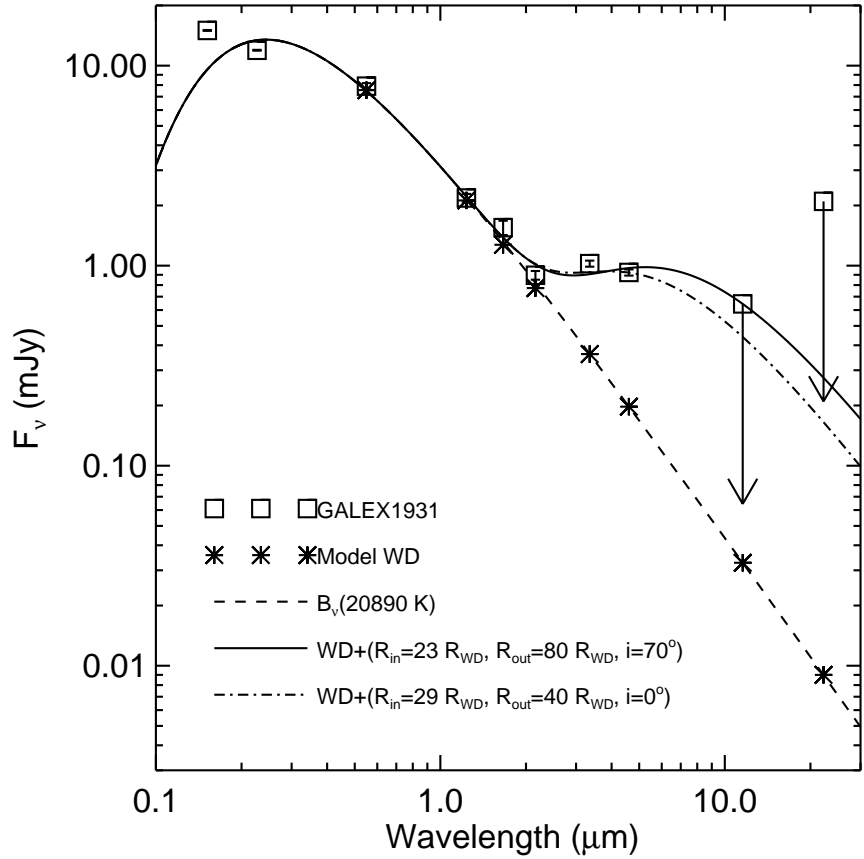}
\caption{\label{fig:f2} Photometry of GALEX1931 compared to synthetic photometry of a similar WD using the models of \citet{bergeron95}, a black body with $T_{eff}$=20890 and scaled to the V and J photometry of GALEX1931, and two different dusty disk models assuming a flat optically thick disk.}
\end{figure}

\begin{figure}
\plotone{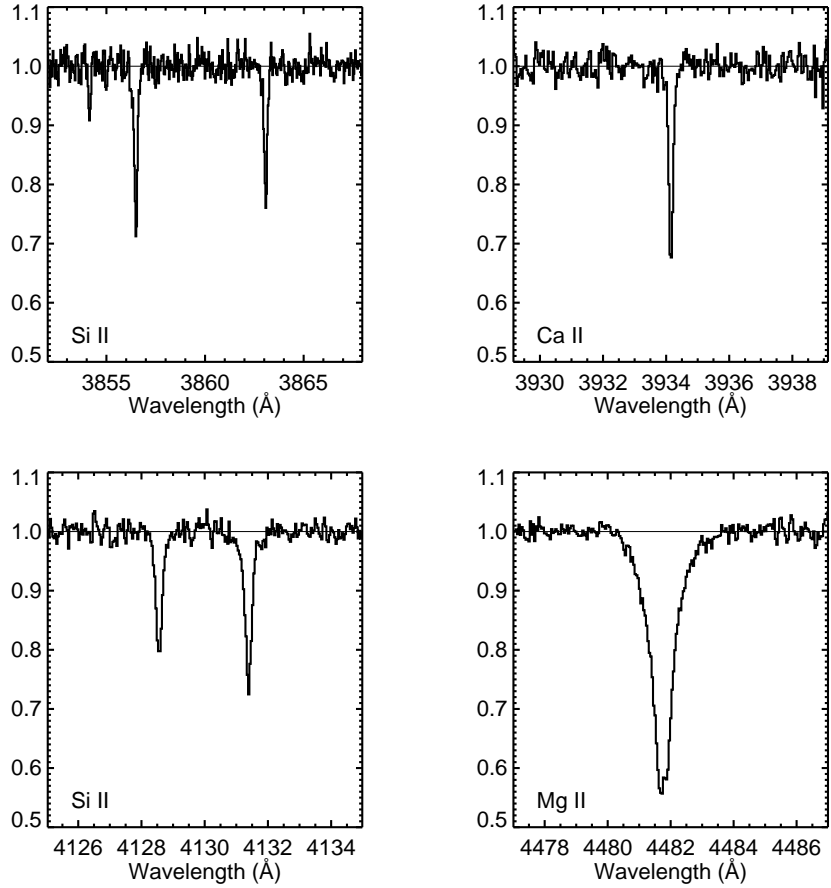}
\caption{\label{fig:f3} Selected regions of our high S/N combination spectrum of GALEX1931 in the optical using the MIKE spectrograph on the Magellan Telescopes, including the region around Mg II (4481\AA), Ca II K (3933\AA), and several Si II lines (3853\AA, 3856\AA, 3862\AA, 4128\AA, 4131\AA).}
\end{figure}

\end{document}